\documentclass{article}
\usepackage{amsmath,amssymb,amsthm}

\newcommand{\N}{{\mathbb N}}

\newcommand{\R}{{\mathbb R}}
\newcommand{\Z}{{\mathbb Z}}
\newcommand{\C}{{\mathbb C}}

\begin{document}
\bibliographystyle{plain}

\centerline{\LARGE{\bf Generalized Heine Identity for}}
\vskip 0.1 truecm
\centerline{\LARGE{\bf Complex Fourier Series of Binomials}}
\vskip 0.5 truecm
\centerline{\Large{Howard S. Cohl$^{\textrm{a}\ast}$, 
Diego E. Dominici$^{\textrm{b}\dag}$}}
\vskip 0.4 truecm
\normalsize{$^{\textrm{a}}$Department of Mathematics, University of Auckland, 
38 Princes Street, Auckland, New Zealand}
\vskip 0.1 truecm
\normalsize{$^{\textrm{b}}$Department of Mathematics, State University of New York, 
New Paltz, New Paltz, New York, U.S.A.}

\vspace{0.3cm}


\begin{center}
\begin{minipage}{0.7\textwidth}
  In this paper we generalize an identity first given by Heinrich Eduard Heine 
  in his treatise, {\it Handbuch der Kugelfunctionen, Theorie und Anwendungen} 
  (1881), which gives a Fourier series for 
  $1/[z-\cos\psi]^{1/2}$, for $z,\psi\in\R$, and  $z>1$, in terms of associated
  Legendre functions of the second kind with odd-half-integer degree and vanishing
  order.  In this paper we 
  give a generalization of this identity as a Fourier series of $1/[z-\cos\psi]^\mu$, 
  where $z,\mu\in\C$, $|z|>1$, and the coefficients of the
  expansion are given in terms of the same functions with order given by $\frac12-\mu$.
  We are also able to compute certain
  closed-form expressions for associated Legendre functions of the second kind.
\end{minipage}
\end{center}
\vspace{-0.2cm}
\begin{center}
\begin{minipage}{0.7\textwidth}
\end{minipage}
\end{center}
\vspace{-0.2cm}
\begin{center}
\begin{minipage}{0.7\textwidth}
\end{minipage}
\end{center}

\vspace{0.2cm}
\noindent $\ast$E-mail:~h.cohl@math.auckland.ac.nz\\
\noindent $\dag$E-mail:~dominicd@newpaltz.edu

\newpage

\section{Introduction}

In this paper we derive and give examples of a generalized Fourier series for 
complex binomials of the form $[z-\cos\psi]^{-\mu}$, where $\psi\in\R$, $\mu\in\C,$ 
$z\in\C\setminus(-\infty,1]$ and $|z|>1$.  We show that the Fourier series is given 
compactly with coefficients given in terms of odd-half-integer 
degree, general complex-order associated Legendre functions of the second kind.  
(Hereafter we refer to associated Legendre functions simply as Legendre functions.)
Algebraic functions such as this naturally arise in classical physics through 
the theory of fundamental solutions of Laplace's equation, and they represent powers of
distances between two points in a  Euclidean geometry.

Fourier expansions for algebraic distance functions have a rich history,
and this expansion makes its appearance in the 
theory of arbitrarily-shaped charge distributions in electrostatics 
(\cite{Pop}, \cite{Walt}, \cite{Barlow}, \cite{PustoI}, \cite{PustoII}, \cite{Verdu}),
magnetostatics
(\cite{Selv}, \cite{Beleggia})
quantum direct and exchange Coulomb interactions
(\cite{CRTB}, \cite{Gatto}, \cite{Enriq}, \cite{Poddar}, \cite{Bagheri}),
Newtonian gravity
(\cite{EvenToh}, \cite{Schachar}, \cite{HureI}, \cite{HureII}, \cite{Fromang}, \cite{Ou},
\cite{Saha}, \cite{Chan}, \cite{Mellon}, \cite{Boley}, \cite{Selv1}), the Laplace coefficients 
for planetary disturbing function
(\cite{DEliseoI}, \cite{DEliseoII}),
and potential fluid flow around actuator discs 
(\cite{BrsAnd}, \cite{HoughOrdway}), 
just to name a few direct physical applications.  A precise Fourier
$e^{im\phi}$ analysis for these applications is extremely useful to fully describe the
general non-axisymmetric nature of these problems.

The fact that this classical theory becomes {\em recently} relevant is due to 
several reasons.  Firstly the fundamental solutions for Laplace's equation are 
ubiquitous in Pure and Applied Mathematics, Physics, and Engineering. 
Secondly the fact that Fourier expansions encapsulate rotationally-invariant symmetries
of geometrical shapes makes it an ideal model case for critical study above and beyond 
the purely spherically symmetric shape, and therefore it provides a powerful tool 
when implemented numerically in a variety of important problems.   Thirdly, Fourier 
expansions for fundamental solutions gives rise to a hardly-studied aspect of Special 
Function Theory, which allows one to further explore the properties of 
higher transcendental functions.

\section{Generalized Heine identity}

Gauss (1812) 
(\cite{Gauss}, p.~128 in Werke III) was able 
to write down closed-form expressions for the 
Fourier series for the related function $[r_1^2+r_2^2-2r_1r_2\cos\psi]^{-\mu}$.
Gauss recognised that the coefficients of the expansion are given in 
terms of the$\ _2F_1$ hypergeometric function, and he was able to write down
a closed-form solution (where we have used modern notations) given by
\begin{equation}
\frac{1}{[r_1^2+r_2^2-2r_1r_2\cos\psi]^\mu}
=\sum_{n=0}^\infty
\epsilon_n\cos(n\psi) \frac{(\mu)_n}{n!}\frac{r_2^n}{r_1^{2\mu+n}}
\ _2F_1\left( n+\mu, \mu; n+1; \frac{r_2^2}{r_1^2}\right),
\label{gaussorig}
\end{equation}

\noindent assuming $r_1,r_2\in\R$, $r_2<r_1$, $\epsilon_n$ is the Neumann factor 
\cite{MorseFesh}
$\epsilon_n=2-\delta_{n,0}$ where $\delta_{n,0}$ is the Kronecker delta, commonly 
occurring in Fourier expansions, $(\mu)_n$ is a Pochhammer symbol for a rising factorial
(see eqs.~(\ref{pochrise}) and (\ref{gamrise})), and the Gauss hypergeometric function 
is defined through
\[
\ _2F_1(a,b;c;z)=\sum_{k=0}^\infty \frac{(a)_k(b)_k}{(c)_kk!}z^k.
\]
\noindent Carl Neumann (1864) (\cite{Neum}, p. 33-34 therein) was one of the first to 
study separable solutions to Laplace's equation in toroidal coordinates.  He examined and wrote 
down the Fourier expansion in term of Gauss hypergeometric functions for the instance 
$\mu=\frac12$.  

But it wasn't until Heine (1881) in his {\it Handbuch der Kugelfunctionen} 
(\cite{Heine}, p.286 therein) that it was recognised that this particular 
Gauss hypergeometric function represented a certain special class of higher 
transcendental functions, namely Legendre functions of the second kind with 
odd-half-integer degree.  These Legendre functions, and in particular, those 
with integer order, are toroidal harmonics, the functions which separate 
Laplace's equation in toroidal coordinates.  The fact that the algebraic 
function of present study relates to behaviour on toroids of revolution is 
an important consequence of a non-axisymmetric Fourier description of potential 
theory in rotationally-invariant coordinate systems.  We now proceed to build 
upon Heine's formula in order to derive a generalization of his identity.

\subsection{Derivation of the identity}

We are interested in computing the following Fourier expansion 

\begin{equation}
\frac{1}{[z-\cos\psi]^\mu}=\sum_{n=0}^\infty \cos(n\psi) A_{\mu,n}(z),
\label{Fourev}
\end{equation}

\noindent where $\mu\in\C$, $z\in\C\setminus(-\infty,1]$, and $|z|>1$.

The expression for these Fourier coefficients is given in the standard manner by

\[
A_{\mu,n}(z)=\frac{\epsilon_n}{\pi}\int_0^\pi \frac{\cos(n\psi)}{[z-\cos\psi]^\mu} d\psi.
\]

\noindent  We can collapse to purely non-negative modes if the 
summand is an even function of the Fourier quantum number.  In that case
\[
\displaystyle \sum_{n=-\infty}^\infty A_{\mu,n}(z) 
e^{in\psi} =
\sum_{n=0}^\infty \epsilon_n A_{\mu,n}(z) \cos(n\psi).
\]

In order to do this, we must show that
\begin{equation}
\arg\left(z-\cos\psi\right)=\arg(z)+\arg\left(1-\frac{\cos\psi}{z} \right).
\label{arg}
\end{equation}
Eq.~(\ref{arg}) is verified as follows.  We define the 
function $f:[-1,1]\rightarrow\R$ such that
\[
f(x)=\arg\left(z+x\right)-\arg(z)-\arg\left(1+\frac{x}{z} \right).
\]
$f$ is clearly continuous and since $[-1,1]$ is connected, $f([-1,1])$ must be 
connected and $f([-1,1])\subset 2\pi\Z$.  Hence $f([-1,1])$ is a one-point set
and since $f(0)=0$, $f$ is a constant equal to zero. 
Therefore we have shown that eq.~(\ref{arg}) 
is true and hence we can rewrite the left-hand side of eq.~(\ref{Fourev}) 
without loss of generality as
\[
{\displaystyle \frac{1}{\left[z-\cos\psi\right]^{\mu}}=
\frac{1}{z^\mu\left[1-{\displaystyle \frac{\cos\psi}{z}}\right]^{\mu}}.}
\]
\noindent Since 
$|z|>1$ and 
$\cos\psi\in [-1,1]$ this implies that $|(\cos\psi)/z|<1$, and we are in a 
position to employ Newton's binomial series
\begin{equation}
(1+w)^\mu=\sum_{k=0}^\infty 
\binom{\mu}{k}
w^k
\label{sbs}
\end{equation}
\noindent where $w,\mu\in\C$, $|w|<1$, and 
\[
\binom{\mu}{k}=\frac{1}{k!}\frac{\Gamma(\mu+1)}{\Gamma(\mu-k+1)}
\]
\noindent is the generalized binomial coefficient.

In what follows we utilize Pochhammer symbols representing either rising or 
falling factorials.  Unfortunately there is no standard convention used for 
Pochhammer symbols.  So here we utilize the following convention which is 
consistent with usage in Special Function theory (i.e. with hypergeometric 
functions).  The Pochhammer symbol for rising factorial is given by

\begin{equation}
(z)_n=
\left\{ \begin{array}{ll}
\displaystyle 1, & \mathrm{if} \qquad n=0;\\[5pt]
\displaystyle (z)\cdot (z+1)\cdot (z+2)\cdots (z+n-1) , & \mathrm{if} \qquad  n\ge 1, 
\end{array} \right. 
\label{pochrise}
\end{equation}

\noindent where $n\in\N_0=\N\cup\{0\}$ and $z\in\C.$  Note that the Pochhammer symbol for the
rising factorial is expressible in terms of Gamma functions as
\begin{equation}
(z)_n=\frac{\Gamma(z+n)}{\Gamma(z)}.
\label{gamrise}
\end{equation}
Similarly the Pochhammer symbol for 
the falling factorial, with the same quantities, is given by
\[
[z]_n=
\left\{ \begin{array}{ll}
\displaystyle 1, & \mathrm{if} \qquad n=0;\\[5pt]
\displaystyle (z)\cdot (z-1)\cdot (z-2)\cdots (z-n+1) , & \mathrm{if} \qquad  n\ge 1. \nonumber
\end{array} \right. 
\]
The Pochhammer symbol for the falling factorial is also expressible in terms of Gamma functions as
\[
[z]_n=\frac{\Gamma(z+1)}{\Gamma(z-n+1)}.
\]

\noindent One can clearly relate the two symbols as such
\begin{equation}
(-1)^n[-z]_n=(z)_n.
\label{pochiden}
\end{equation}

\noindent The binomial coefficients can be defined through the Pochhammer symbol for
the falling factorial
\[
\binom{z}{n}=\frac{[z]_n}{n!}.
\]
Combining this relation for the binomial coefficients, eq.~(\ref{pochiden}), and eq.~(\ref{sbs}),
we obtain

\begin{equation}
\frac{1}{[z-\cos\psi]^\mu}=\sum_{k=0}^\infty \frac{(\mu)_k}{k!} z^{-\mu-k}\cos^k\psi.
\label{Fourev2}
\end{equation}

\noindent We can expand the powers of cosine using the following trigonometric identity

\[
\cos^k\psi=\frac{1}{2^k}\sum_{n=0}^k
\binom{k}{n} \cos[(2n-k)\psi],
\]
\noindent which is the standard expansion for powers using Chebyshev polynomials (see for instance
p.52 in Fox and Parker (1968) \cite{FoxParker}).
Inserting this expression in eq.~(\ref{Fourev2}), we obtain the following double summation
expression
\begin{equation}
\frac{1}{[z-\cos\psi]^\mu}=
\sum_{k=0}^\infty \sum_{n=0}^k\frac{(\mu)_k}{k!}\frac{1}{2^kz^{\mu+k}}
\binom{k}{n}
\cos[(2n-k)\psi].
\label{dbsum}
\end{equation}

Now we perform a double index replacement in eq.~(\ref{dbsum}).  There are two 
separate cases $k\le 2n$ and $k\ge 2n$.  There is an overlap if both have an equality and in that case we must multiply by $1/2$ after
we sum over both cases.  If $k\le2n$ make the substitution 
$k^\prime=k-n$ and $n^\prime=2n-k$.  It follows
that $k=2k^\prime+n^\prime$ and $n=n^\prime+k^\prime$, therefore
\[
\binom{k}{n}=\binom{2k^\prime+n^\prime}{n^\prime+k^\prime}=
\binom{2k^\prime+n^\prime}{k^\prime}.
\]

\noindent If $k\ge 2n$ make the substitution $k^\prime=n$ and $n^\prime=k-2n$.  
Then 
$k= 2k^\prime + n^\prime$  and $n=k^\prime$ therefore
\[
\binom{k}{n}=
\binom{2k^\prime+ n^\prime}{k^\prime}=
\binom{2k^\prime+n^\prime}{k^\prime+n^\prime},
\]

\noindent where the equalities of the binomial coefficients are confirmed using
the identity 
\begin{equation}
\binom{l}{m}=\binom{l}{l-m},
\label{biniden}
\end{equation}
where $l\in\Z$.  To take into account the double counting which occurs when $k=2n$
(which occurs when $n^\prime=0$) we introduce a factor of $\epsilon_{n^\prime}/2$ 
into the expression
(and relabel $k^\prime\mapsto k$ and $n^\prime\mapsto n$).  We are left with
\begin{eqnarray}
\frac{1}{[z-\cos\psi]^\mu}=
&{\displaystyle \frac{1}{2z^\mu}}& \sum_{n=0}^\infty \epsilon_n \cos(n\psi)\nonumber\\[0.2cm]
&\times&\sum_{k=0}^\infty 
\frac{(\mu)_{2k+n}}{(2k+n)!}\frac{1}{(2z)^{2k+n}}
\left[
\binom{2k+n}{k} + \binom{2k+n}{k+n}
\right],\nonumber
\end{eqnarray}
which is straightforwardly simplified using the definition of the 
binomial coefficients and eq.~(\ref{biniden})
\[
\frac{1}{[z-\cos\psi]^\mu}=
\frac{1}{z^\mu} \sum_{n=0}^\infty \epsilon_n \cos(n\psi)
\frac{1}{(2z)^n}
\sum_{k=0}^\infty 
\frac{(\mu)_{2k+n}}{k!(k+n)!}
\frac{1}{4^k}
\left(
\frac{1}{z^2}
\right)^k.
\]

\noindent The second sum is given in terms of a Gauss hypergeometric series

\begin{equation}
\frac{1}{[z-\cos\psi]^\mu}=
\sum_{n=0}^\infty \epsilon_n \cos(n\psi)
\frac{(\mu)_n}{n!2^nz^{\mu+n}}
\ _2F_1
\left(
\frac{\mu+n}{2},
\frac{\mu+n+1}{2};
n+1;
\frac{1}{z^2}
\right).
\label{finsum}
\end{equation}

\noindent This is just a re-statement of Gauss's original formula, eq.~(\ref{gaussorig}),
with 
\[
z=\frac{r_1^2+r_2^2}{2r_1r_2},
\]
only now the quantities $z,\mu\in\C.$  This Gauss hypergeometric function is expressible in terms of
the Legendre function of the second kind 
(Chapter 8 by I.~A.~Stegun \cite{Abra}, p.332)

\begin{eqnarray}
&\displaystyle \!\ _2F_1\left(\frac{\mu+\nu}{2},\frac{\mu+\nu+1}{2};\nu+1;\frac{1}{z^2}\right)&\nonumber\\[0.2cm]
&\displaystyle=\sqrt{\frac{2}{\pi}}
\frac{ 2^\nu\Gamma(\nu+1) z^{\nu+\mu} (z^2-1)^{-\mu/2+1/4}e^{-i\pi(\mu-1/2)}}
{\Gamma(\nu+\mu)}
 Q_{\nu-1/2}^{\mu-1/2}(z),&
\label{gaussQ}
\end{eqnarray}

\noindent where $\nu,\mu\in\C$, a branch cut is chosen along the
real axis along $(-\infty,1],$ and otherwise this expression is valid 
except where the constituent functions are not defined.

If we substitute $\nu=n\in\Z$ in the hypergeometric function and take
advantage of the following property of Legendre functions of the second
kind with $|z|>1$ \cite{CTRS} 
\[
Q_{-n-1/2}^\mu(z)=Q_{n-1/2}^\mu(z),
\]
in eq.~(\ref{finsum}), for all $n\in\Z$ and $\mu\in\C$, we obtain a complex generalization of Heine's 
reciprocal square root identity
given as follows 
\begin{equation}
\displaystyle
\frac{1}{[z-\cos\psi]^\mu}=
\sqrt{\frac{2}{\pi}}\frac{e^{-i\pi(\mu-1/2)}(z^2-1)^{-\mu/2+1/4}}{\Gamma(\mu)}
\sum_{n=-\infty}^{\infty}e^{in\psi} Q_{n-1/2}^{\mu-1/2}(z),
\label{compgen}
\end{equation}
\noindent where Heine's original identity (\cite{Heine}, p.286)
is the case for $\mu=\frac12$ given by
\begin{equation}
\frac{1}{\sqrt{z-\cos\psi}}=\frac{\sqrt{2}}{\pi}
\sum_{m=-\infty}^{\infty} Q_{m-\frac{1}{2}}(z)\ e^{im\psi},
\label{Heine}
\end{equation}
see \cite{CT} for exact forms and recurrence relations for these Legendre functions.
Eq.~(\ref{Heine}) was recently generalized for $\mu$ 
given by odd-half-integers in Selvaggi et al. (2008) (\cite{Selv}, p.~033913-6).  

The generalization given by eq.~(\ref{compgen}) is also expressible in terms of 
Legendre functions of the first kind through Whipple's transformation of Legendre
functions (\cite{CTRS}, \cite{Abra}) as
,,
\[
\frac{1}{[z-\cos\psi]^\mu}=
\frac{(z^2-1)^{-\mu/2}}{\Gamma(\mu)}
\sum_{n=-\infty}^{\infty}e^{in\psi} \Gamma(\mu-n)
P_{\mu-1}^n\left(\frac{z}{\sqrt{z^2-1}}\right).
\]

In the following section we describe some specific examples of the generalized
identity, and present some interesting implications.

\subsection{Examples and implications}

We now have the value of the following definite integral
\[
\int_{-\pi}^\pi \frac{\cos(nt)dt}{[z-\cos t]^\mu}=
2^{3/2}\sqrt{\pi}\frac{e^{-i\pi(\mu-1/2)}(z^2-1)^{-\mu/2+1/4}}{\Gamma(\mu)}
Q_{n-1/2}^{\mu-1/2}(z).
\]

The principal example for the generalized Heine identity which was
first proven in Selvaggi et al. (2008) \cite{Selv} (see also \cite{MOS}, p.~182 and \cite{CPortent})
 for $\mu=q+\frac12$ where $q\in\Z$ 
is given by
\[
\frac{1}{[z-\cos\psi]^{q+1/2}}=
\frac{2^{q+1/2}(-1)^q}{\pi(2q-1)!!(z^2-1)^{q/2}}
\sum_{n=0}^\infty \epsilon_n \cos(n\psi) Q_{n-1/2}^q(z),
\]
where $z!!$ is the double factorial (Chapter 6 by P.~J.~Davis \cite{Abra}).  Notice that the double factorial is 
well-defined for negative odd integers \cite{Arf}
\[
(-2q-1)!!=\frac{(-1)^n\ 2^n\ n!}{(2n)!}.
\]
For instance, for $q=-1$ we have
\[
\sqrt{z-\cos\psi}=\frac{\sqrt{z^2-1}}{\sqrt{2}\pi}
\sum_{n=0}^\infty \frac{\epsilon_n \cos(n\psi)}{n^2-\frac14} Q_{n-1/2}^1(z),
\]
and for $q=+1$ 
\begin{equation}
[z-\cos\psi]^{3/2}=\frac{-2^{3/2}}{\pi\sqrt{z^2-1}}
\sum_{n=0}^\infty \epsilon_n \cos(n\psi) Q_{n-1/2}^1(z).
\label{zthreeh}
\end{equation}
Note that the minus sign in eq.~(\ref{zthreeh}) expansion might seem initially 
troublesome, except that it is important to notice that the unit-order Legendre functions
of the second kind are all negative in sign as can be easily seen from the Gauss 
hypergeometric function representation, eq.~(\ref{gaussQ}).  We have
as such
\[
Q_{-1/2}^1(z)=\frac{-1}{\sqrt{2(z-1)}}E\left(\sqrt{\frac{2}{z+1}}\right),
\]
\noindent and
\[
Q_{1/2}^1(z)=\frac{-z}{\sqrt{2(z-1)}}E\left(\sqrt{\frac{2}{z+1}} \right) 
+\sqrt{\frac{z-1}{2}}K\left(\sqrt{\frac{2}{z+1}} \right),
\]
where $K$ and $E$ are the complete elliptic integrals of the first and second kind respectively.  
The rest of the unit-order, odd-half-integer degree Legendre functions of the
second kind can be computed using the following recurrence relation
\[
Q_{m+1/2}^1(z)=\frac{4mz}{2m-1}Q_{m-1/2}^1(z)-\frac{2m+1}{2m-1}Q_{m-3/2}^1(z).
\]
Notice that all 
odd-half-integer degree, integer order Legendre functions can be written as 
linear combinations of elliptic integrals of the first and second kind.  The analogous
formulas for $q=0$ are given in Cohl and Tohline (1999) \cite{CT}, eqs. (22)--(26).

Let us look at its behaviour for $\mu$ being a negative integer such that the
binomial expansion should reduce to a polynomial in $z$.  Let us take the
limit as $\mu\rightarrow -q$.  Using the negative-order condition for Legendre functions 
of the second kind (\cite{CTRS}, p.367)
\[
Q_{p-1/2}^{-n-1/2}(z)=e^{-i\pi}\frac{\Gamma(p-n)}{\Gamma(p+n+1)}Q_{p-1/2}^{n+1/2}(z),
\]

\noindent and
\begin{equation}
(z-\cos\psi)^q=-i\sqrt{\frac{2}{\pi}}(-1)^q(z^2-1)^{q/2+1/4}
\sum_{n=0}^\infty
\epsilon_n\cos(n\psi)
\frac{(-q)_n}{(q+n)!}
Q_{n-1/2}^{q+1/2}(z),
\label{integer}
\end{equation}

\noindent where we have used eq.~(\ref{gamrise}). Let us explicitly verify 

\[
L_0=\lim_{q\rightarrow 0}(z-\cos\psi)^q=1.
\]

\noindent Evaluating the limit of the right-hand side of eq.~(\ref{integer}) yields

\[
L_0=\sqrt{\frac{2}{\pi}}
\frac{(z^2-1)^{1/4}}{i}
\sum_{n=0}^\infty
\frac{\epsilon_n\cos(n\psi)}{n!}
Q_{n-1/2}^{1/2}(z)
\lim_{q\rightarrow 0}
\frac{\Gamma(n-q)}{\Gamma(-q)},
\]
\noindent where we have used
\[
\lim_{q\rightarrow p} \frac{\Gamma(n-q)}{\Gamma(-q)}=(-p)_n.
\]
Hence
for $p=0$, only the first term in the sum survives.  The Legendre function
\[
Q_{-1/2}^{1/2}(z)=i\sqrt{\frac{\pi}{2}}(z^2-1)^{-1/4},
\]

\noindent is given in Abramowitz \& Stegun (1964) (\cite{Abra}, p.334),
reduces the limit to 1, as expected.  Take for instance 

\[
L_1=\lim_{q\rightarrow 1}(z-\cos\psi)^q=z-\cos\psi.
\]

\noindent Evaluating the limit of the right-hand side 
of eq.~(\ref{integer}) yields

\[
L_1=\sqrt{\frac{2}{\pi}}
i(z^2-1)^{3/4}
\sum_{n=0}^\infty
\frac{\epsilon_n\cos(n\psi)}{(n+1)!}
Q_{n-1/2}^{3/2}(z)
\lim_{q\rightarrow 1}
\frac{\Gamma(n-q)}{\Gamma(-q)}.
\]

\noindent The limit in the sum is given by 1 for $n=0, -1$ for $n=1$ and
zero otherwise, hence only the first two terms in the sum survive.  
Using the Legendre functions

\begin{eqnarray}
Q_{-1/2}^{3/2}(z)=\frac{1}{i}\sqrt{\frac{\pi}{2}}z(z^2-1)^{-3/4},\nonumber\\[0.2cm]
Q_{1/2}^{3/2}(z)=\frac{1}{i}\sqrt{\frac{\pi}{2}}(z^2-1)^{-3/4},\nonumber
\end{eqnarray}

\noindent we arrive at the desired result for $L_1$.

Substituting $\mu=q\in\N$ in eq.~(\ref{compgen}) yields
\begin{equation}
\frac{1}{[z-\cos\psi]^q}=
i\sqrt{\frac{2}{\pi}}\frac{(-1)^q(z^2-1)^{-q/2+1/4}}{(q-1)!}
\sum_{n=0}^\infty\epsilon_n \cos(n\psi) Q_{n-1/2}^{q-1/2}(z),
\label{genint}
\end{equation}
which the right-hand side is verified to be positive real upon examination
of the definition of the Legendre function of the second kind in term of
the Gauss hypergeometric function, eq.~(\ref{gaussQ}).  For $q=1$ the
coefficients are Legendre functions with odd-half-integer degree and $\frac12$ order.
These can be evaluated using (\cite{Abra}, eq.~(8.6.10))
\begin{equation}
Q_\nu^{1/2}(z)=i\sqrt{\frac{\pi}{2}}(z^2-1)^{-1/4}\left[z+\sqrt{z^2-1} \right]^{-\nu-1/2},
\label{qnuh}
\end{equation}
valid for $z,\nu\in\C$ and $|z|>1$.  Note we also have (\cite{Abra}, eq.~(8.6.11))
\begin{equation}
Q_\nu^{-1/2}(z)=-i\frac{\sqrt{2\pi}}{2\nu+1}(z^2-1)^{-1/4}\left[z+\sqrt{z^2-1} \right]^{-\nu-1/2}.
\label{qnumh}
\end{equation}

If we take $z=\cosh\eta$ and $\nu=n-\frac12$, where 
$n\in\Z$ and insert the resulting expression in eq.~(\ref{genint}) we obtain
\begin{equation}
\frac{1}{\cosh\eta-\cos\psi}=\frac{1}{\sinh\eta}\sum_{n=0}^\infty 
\epsilon_n e^{-n\eta} \cos(n\psi).
\label{genint1}
\end{equation}
Similarly if we use eq.~(\ref{qnumh})
and the order recurrence relation for Legendre functions
\[
Q_\nu^{\mu+2}(z)=-2(\mu+1)\frac{z}{\sqrt{z^2-1}}Q_\nu^{\mu+1}(z)+(\nu-\mu)(\nu+\mu+1)Q_\nu^\mu(z),
\]
we are able to compute all required odd-half-integer order Legendre functions appearing in 
eq.~(\ref{genint}).  

By taking $z=\cosh\eta$ we have 
\begin{eqnarray}
&\displaystyle \frac{1}{(\cosh\eta-\cos\psi)^2}=
\frac{1}{\sinh^3\eta}
\sum_{n=0}^\infty
\epsilon_n
\cos(n\psi)
e^{-n\eta}
(n\sinh\eta+\cosh\eta),&\nonumber\\[0.2cm]
&\displaystyle\frac{1}{(\cosh\eta-\cos\psi)^3}=
\frac{1}{2\sinh^5\eta}
\sum_{n=0}^\infty
\epsilon_n
\cos(n\psi)
e^{-n\eta}&\nonumber\\[0.2cm]
&\times\left((n^2-1)\sinh^2\eta
+3n\sinh\eta\cosh\eta+3\cosh^2\eta\right),\nonumber
\end{eqnarray}
\noindent and
\begin{eqnarray}
&\displaystyle\frac{1}{(\cosh\eta-\cos\psi)^4}=
\frac{1}{6\sinh^7\eta}
\sum_{n=0}^\infty
\epsilon_n
\cos(n\psi)
e^{-n\eta}&\nonumber\\[0.2cm]
&\displaystyle\times\left((n^3-4n)\sinh^3\eta+(6n^2-9)\sinh^2\eta\cosh\eta
+15n\sinh\eta\cosh^2\eta+15\cosh^3\eta\right).&\nonumber
\end{eqnarray}

One way to verify these formulae is to start with the the generating 
function for Chebyshev polynomials of the first kind $T_n(x)$ (\cite{MOS} and \cite{FoxParker}, p.51)
\[
\frac{1-z^2}{1+z^2-2xz}=\sum_{n=0}^\infty \epsilon_n z^n T_n(x),
\]
and substitute $z=\cosh\eta$, yielding eq.~(\ref{genint1}).  The rest of the examples can 
be verified by direct repeated differentiation with respect to $\eta$.

\section{Closed-form expressions for certain Legendre functions}

We have%
\begin{eqnarray}
A_{\mu,n}(z)&=&\frac{(\mu)_n}{2^{n}n!z^{\mu+n}}%
\ _{2}F_{1}\left(  \frac{\mu+n}{2},\frac{\mu+n+1}{2};n+1;z^{-2}\right)\nonumber\\[0.2cm]
&=&\sqrt{\frac{2}{\pi}}\frac{e^{-i\pi(\mu-1/2)}(z^2-1)^{-\mu/2+1/4}}{\Gamma(\mu)}
Q_{n-1/2}^{\mu-1/2}(z).
\label{aqeq}
\end{eqnarray}

\noindent Let
\[
z=\cosh\left(  \eta\right)  =\frac12\left(x+x^{-1}\right),
\]
with
\[
x=e^{-\eta}.
\]
Since $\eta>0,$ we have $0<x<1,$ and
\begin{eqnarray}
A_{\mu,n}\left(  \cosh\eta  \right)  =\frac{\left(  \mu\right)
_{n}}{n!}2^{\mu}x^{\mu+n}\left(  1+x^{2}\right)  ^{-\left(  \mu+n\right)
}\nonumber\\[0.2cm]
\times\ _{2}F_{1}\left(  \frac{\mu+n}{2},\frac{\mu+n+1}{2};n+1;\frac{4x^{2}%
}{(1+x^{2})^2}\right).
\end{eqnarray}
Using Andrews, Askey, \& Roy (1999) (\cite{AAR} eq.~(3.1.9)) with $a=\mu+n$ and $b=\mu,$ we get%
\[
A_{\mu,n}\left(  \cosh \eta  \right)  =\frac{\left(  \mu\right)
_{n}}{n!}2^{\mu}x^{\mu+n}\ _{2}F_{1}\left(  \mu+n,\mu;n+1;x^{2}\right)  .
\]

\noindent Using Pfaff's transformation
(\cite{AAR}, eq.~(2.2.6)), we obtain%
\begin{align*}
A_{\mu,n}\left(  \cosh\eta \right)    & =\frac{\left(
\mu\right)  _{n}}{n!}2^{\mu}x^{\mu+n}\ _{2}F_{1}\left(  \mu,\mu+n;n+1;x^{2}%
\right)  \\
& =\frac{\left(  \mu\right)  _{n}}{n!}2^{\mu}x^{\mu+n}\left(  1-x^{2}\right)
^{-\mu}\ _{2}F_{1}\left(  \mu,1-\mu;n+1;\frac{x^{2}}{x^{2}-1}\right)  .
\end{align*}
Taking $\mu=q\in\N$, we can write%
\[
A_{q,n}\left(  \cosh\eta  \right)  =\frac{\left(  q\right)
_{n}}{n!}2^{q}\frac{x^{q+n}}{\left(  1-x^{2}\right)  ^{q}}\sum
_{k=0}^{q-1}\frac{\left(  q\right)  _{k}\left(  1-q\right)  _{k}%
}{\left(  n+1\right)  _{k}}\frac{1}{k!}\left(  \frac{x^{2}}{x^{2}-1}\right)
^{k}%
\]
or%
\begin{align*}
A_{q,n}\left(\cosh\eta\right)    & =2^{q}\frac
{x^{q+n}}{\left(  1-x^{2}\right)  ^{q}}\sum_{k=0}^{q-1}\binom{q
+n-1}{n+k}\binom{q+k-1}{k}\left(  -1\right)  ^{k}\left(  \frac{x^{2}}%
{x^{2}-1}\right)  ^{k}\\
& =\frac{x^{n}}{\left(  \frac{1-x^{2}}{2x}\right)  ^{q}}\sum_{k=0}^{q
-1}\binom{q+n-1}{n+k}\binom{q+k-1}{k}2^{-k}{\displaystyle \left(  \frac{x}{\frac{1-x^{2}%
}{2x}}\right)  ^{k}}.
\end{align*}
Since $x=e^{-\eta},$ we get
\[
A_{q,n}\left(  \cosh\eta  \right)  =\frac{e^{-n\eta}}%
{\sinh^{q}\left(  \eta\right)  }\sum_{k=0}^{q-1}\binom{q+n-1}{n+k}%
\binom{q+k-1}{k}\frac{e^{-k\eta}}{2^k\sinh^{k}\eta},
\]

\noindent or for instance by using eq.(\ref{biniden})
\[
A_{q,n}(\cosh\eta)=\frac{e^{-n\eta}}{\sinh^q(\eta)}
\sum_{k=0}^{q-1}
\binom{n+q-1}{q-k-1} \binom{q+k-1}{q-1}
\frac{e^{-k\eta}}{2^k\sinh^k\eta}.
\]
Using this formula and eq.~(\ref{aqeq}), we are able to write down a concise formula for 
odd-half-integer degree, odd-half-integer order Legendre functions 
\begin{eqnarray}
Q_{n-1/2}^{q-1/2}(z)=\sqrt{\frac{\pi}{2}}
\frac{i(-1)^{q+1}(z-\sqrt{z^2-1})^n}{(q-1)!(z^2-1)^{1/4}}\nonumber\\[0.2cm]
\times\sum_{k=0}^{q-1}
\binom{n+q-1}{q-k-1} \binom{q+k-1}{q-1}
\left[\frac{z-\sqrt{z^2-1}}{2(z^2-1)^{1/2}}\right]^k,\nonumber
\end{eqnarray}
or in terms of Pochhammer symbols
\begin{eqnarray}
Q_{n-1/2}^{q-1/2}(z)=i(-1)^{q+1}\sqrt{\frac{\pi}{2}}\frac{\Gamma(q+n)}{n!}
\frac{(z-\sqrt{z^2-1})^n}{(z^2-1)^{1/4}}\nonumber\\[0.2cm]
\times\sum_{k=0}^{q-1}
\frac{(q)_k(1-q)_k}{(n+1)_kk!}
\left[\frac{-z+\sqrt{z^2-1}}{2(z^2-1)^{1/2}}\right]^k,\nonumber
\end{eqnarray}
since 
\[
(-1)^k
\frac{(q)_n(q)_k(1-q)_k}{(n+1)_kk!n!}=
\binom{q+k-1}{k} \binom{q+n-1}{n+k}
.
\]
This leads us to conjecture the following general formula, $q\in\N$
\begin{eqnarray}
Q_\nu^{q-1/2}(z)=
i(-1)^{q+1}\sqrt{\frac{\pi}{2}}
\frac{\Gamma\left(q+\nu+\frac12\right)}{\Gamma\left(\nu+\frac32 \right)}
\frac{(z+\sqrt{z^2-1})^{-\nu-1/2}}{(z^2-1)^{1/4}}\nonumber\\[0.2cm]
\times\sum_{k=0}^{q-1}
\frac{(q)_k(1-q)_k}{\left(\nu+\frac32\right)_kk!}
\left[\frac{-z+\sqrt{z^2-1}}{2(z^2-1)^{1/2}}\right]^k.\nonumber
\end{eqnarray}
This formula is verified using entry.~(31) on p.162 of \cite{MOS}, $q\in\Z$ to obtain
\begin{eqnarray}
Q_\nu^{q-1/2}(z)=
i(-1)^{q+1}\sqrt{\frac{\pi}{2}}
\frac{\Gamma\left(q+\nu+\frac12\right)}{\Gamma\left(\nu+\frac32 \right)}
\frac{(z+\sqrt{z^2-1})^{-\nu-1/2}}{(z^2-1)^{1/4}}\nonumber\\[0.2cm]
\times\sum_{k=0}^{|q-\frac12|-\frac12}
\frac{(q)_k(1-q)_k}{\left(\nu+\frac32\right)_kk!}
\left[\frac{-z+\sqrt{z^2-1}}{2(z^2-1)^{1/2}}\right]^k.
\label{qnuqmhz}
\end{eqnarray}
This is a generalization of the very important general formulae given by
eqs. (8.6.10) and (8.6.11) in \cite{Abra}, p.334.  Taking $z=\cosh\eta$ we have
\begin{eqnarray}
Q_\nu^{q-1/2}(\cosh\eta)=
i(-1)^{q+1}\sqrt{\frac{\pi}{2}}
\frac{\Gamma\left(q+\nu+\frac12\right)}{\Gamma\left(\nu+\frac32 \right)}
\frac{e^{-\eta(\nu+1/2)}}{\sqrt{\sinh\eta}}\nonumber\\[0.2cm]
\times\sum_{k=0}^{|q-\frac12|-\frac12}
\frac{(q)_k(1-q)_k}{\left(\nu+\frac32\right)_kk!}
\frac{(-1)^ke^{-k\eta}}{2^k\sinh^k\eta}.
\label{qnuqmhcosh}
\end{eqnarray}
An alternative procedure for computing these Legendre functions is to start with eqs.~(\ref{qnumh}) 
and (\ref{qnuh}) and use the order recurrence relation, \cite{Grad} eq.~(8.732.3)
\[
Q_\nu^{\mu+2}(z)+\frac{2(\mu+1)z}{\sqrt{z^2-1}}Q_\nu^{\mu+1}(z)
=(\nu-\mu)(\nu+\mu+1)Q_\nu^\mu(z).
\]
On the other hand, the expressions given by eqs.~(\ref{qnuqmhz}) and (\ref{qnuqmhcosh})
directly give closed-form expressions for the Legendre functions simply by evaluating a finite sum.

\section{Conclusion}
Not only is the generalized Heine identity useful for studying Poisson's equation in 
three-dimensions, but it is equally valid with fundamental solutions for $k$ powers of 
the Laplacian in $\R^n.$  These are given in terms of a functional form which matches 
exactly the generalized Heine identity, particularly those in odd dimensions and in the 
even dimensions for $k\le \frac{n}{2}-1$.  
As will be seen in follow-up publications, the generalized Heine identity can be used 
as a powerful tool for expressing geometric properties (multi-summation addition theorems) 
for these fundamental solutions in rotationally-invariant coordinate systems which yield 
solutions to such equations through separation of variables.  

%

%
%

\section{Acknowledgements}
H.~S.~Cohl would like to thank the following people: I had many valuable
discussions with Dr Tom ter Elst and also with Dr Garry J.~Tee who
carefully proofread this article.  The work of D.~Dominici was partially 
supported by a Humboldt Research Fellowship for Experienced Researchers 
from the Alexander von Humboldt Foundation.

\bibliography{refbib}

\end{document}